\documentclass{article}

{}
{}
{}

\usepackage{amsmath} 
\usepackage{amsthm} 
\usepackage{cite} 
\usepackage{hyperref} 
\usepackage{graphics} 
\usepackage{graphicx}
\usepackage{algorithmic} 
\usepackage{hyperref}
\usepackage{authblk}
\usepackage[margin=1.2cm]{geometry}
\usepackage{multicol}
\usepackage{blindtext}
\usepackage{epstopdf}
\usepackage{textcomp}
\usepackage{xcolor}
\usepackage{subcaption}
\usepackage{multirow}
\usepackage{afterpage}
\usepackage{algorithm}
\usepackage[shortlabels]{enumitem}
\usepackage{bm}
\usepackage{tabularx}
\usepackage{makecell}

\begin{document}
\title{RNA-TransCrypt: Image Encryption Using Chaotic RNA Encoding, Novel Transformative Substitution, and Tailored Cryptographic Operations
}

\author[1*]{Muhammad Shahbaz Khan}
\author[1]{Jawad Ahmad}
\author[1]{Ahmed Al-Dubai}
\author[1]{Baraq Ghaleb}
\author[1]{Nikolaos Pitropakis}
\author[1]{William J. Buchanan}

\affil[1]{{School of Computing, Engineering and the Built Environment, Edinburgh Napier University, Edinburgh, UK.} \newline  *Email: MuhammadShahbaz.Khan@napier.ac.uk }
\date{}

\setcounter{Maxaffil}{0}
\renewcommand\Affilfont{\itshape\small}

\maketitle




\begin{abstract}
 Given the security concerns of Internet of Things (IoT) networks and limited computational resources of IoT devices, this paper presents RNA-TransCrypt, a novel image encryption scheme that is not only highly secure but also efficient and lightweight. RNA-TransCrypt integrates the biocryptographic properties of RNA encoding with the non-linearity and unpredictability of chaos theory. This scheme introduces three novel contributions: 1) the two-base RNA encoding method, which transforms the image into RNA strands-like sequence, ensuring efficient scrambling; 2) the transformative substitution technique, which transforms the s-box values before replacing the pixel values, and is responsible for making the scheme lightweight; and 3) three mathematical cryptographic operations designed especially for image encryption that ensure the effective transformation of the s-box values, resulting in a new outcome even for the same input values. These modules are key-dependent, utilizing chaotic keys generated by the De Jong Fractal Map and the Van der Pol Oscillator. Extensive security analysis, including histogram analysis, correlation analysis, and the results of the statistical security parameters obtained from the Gray-Level Co-occurrence Matrix (GLCM) validate the efficacy of the proposed scheme in encrypting input images with close-to-ideal results of 7.997 entropy and 0.0006 correlation. 
\end{abstract}

\begin{multicols}{2}

\section{Introduction}
With the widespread adoption of Internet of Things (IoT) and the increased transmission of digital images across IoT networks, ensuring the security of digital data, especially images, has become a critical concern. As most of the IoT data remains unprotected \cite{neshenko_2019_demystifying}, it poses a significant risk to the integrity and privacy of transmitted image data and, hence, this IoT image data needs protection. Given the limited computational power and energy resources of IoT devices \cite{imteaj_2021_a, dhanda_2020_lightweight}, the image encryption algorithms designed for these devices must be lightweight and efficient, while maintaining high security and efficacy. 

With the recent advancements in biocryptography, RNA encoding, owing to its unique biological properties, has emerged to be a promising technique for image encryption. In addition to DNA encoding \cite{ahmed_2023_a, yang_2022_image}, recently, RNA encoding is extensively being used in encryption algorithms \cite{zhang_2022_an, mahmud_2020_evolutionarybased, lu_2022_exploiting}. Similarly, chaos theory, with its inherent unpredictability and nonlinearity, offers significant potential in the field of image encryption, especially through chaotic maps \cite{daniasaleemmalik_2023_4d, khan2023srss, ali2023cellsecure}. Chaotic maps, in addition to offering a layer of complexity, are also efficient when compared to traditional pseudo-random number generators \cite{abutaha_2022_secure, ahmad_2016_a, }. Chaotic systems, when integrated with lightweight encryption methodologies, enhance the security of the encryption algorithms without imposing computational burden on transmission channels. This is why, the motivation of this paper is to design a simple, lightweight, yet highly secure image encryption scheme for IoT devices, which also takes into account the limited capabilities of IoT nodes, including limited computational power, restricted memory and storage, and constrained network bandwidth.

This paper presents RNA-TransCrypt, a novel image encryption scheme that utilizes RNA encoding, chaos, and a novel transformative substitution method. The core innovation lies in the novel transformative substitution that makes the algorithm lightweight. In addition, three novel cryptographic operations tailored for image encryption have been introduced for the confusion part.\\

\begin{figure*}[!t]
  \centering
  \includegraphics[width=0.8\textwidth]{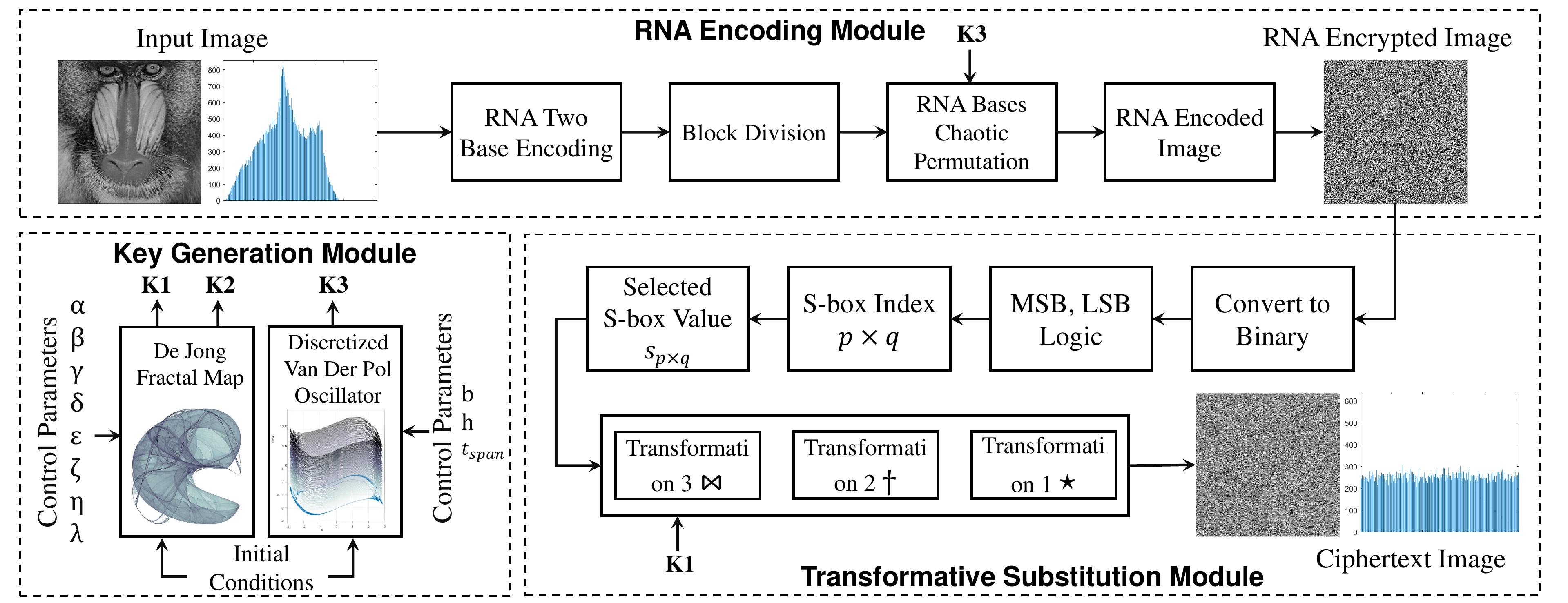}
  \caption{The proposed image encryption scheme---RNA-TransCrypt}
  \label{fig:fig1}
\end{figure*}

The main contributions of this paper are:

\begin{enumerate}
    \item \textbf{Novel Image Encryption Scheme:} This paper presents RNA-TransCrypt, an image encryption scheme that integrates the biocryptographic features of RNA for diffusion/permutation with the transformative substitution method for confusion. Both processes are controlled by chaotic keys generated using the De Jong Fractal Map and the Van der Pol Oscillator.  
    \item \textbf{Two-Base RNA Encoding:} A novel key-dependent two-base RNA encoding scheme is proposed, which transforms the image into a simulated RNA strand-like sequence, ensuring effective scrambling of pixels.
    \item \textbf{Transformative Substitution:} A novel substitution technique, termed transformative substitution, is introduced. This method eliminates the need for multiple s-boxes and substitution rounds, transforming the s-box values into a new value before replacing the pixel in the image.    
    \item \textbf{Cryptographic Mathematical Operations for Image encryption:} Three cryptographic operations tailored for image encryption are designed. These operations include: \( O_1 = (p_{i,j} + s_{p \times q} + K_2) \mod 2^8 \), a key dependent modulus addition operation, \( O_2 = (s_{p \times q} \gg n) \oplus (p_{i,j} \ll (8-n)) \), a bitwise shifting and XOR functions-based operation, and \( O_3 = (\text{MSB}_{1-4}(p_{i,j}) \parallel \text{LSB}_{1-4}(s_{p \times q})) \oplus (\text{MSB}_{5-8}(p_{i,j}) \parallel \text{LSB}_{5-8}(s_{p \times q})) \), a concatenation and XORing of the Most Significant Bits (MSB) and Least Significant Bits (LSB)-based operation. These operations ensure effective transformation of the selected s-box values resulting in additional layer of security.
\end{enumerate}

\section{RNA-TransCrypt---The Proposed Encryption Scheme}
The proposed encryption scheme comprises three important modules as depicted in Figure \ref{fig:fig1}: 1) the 'Key Generation' module; 2) the 'RNA Encoding' module; and 3) the 'Transformative Substitution' module, described in following subsections.
\subsection{The Key Generation Module}
The key generation module utilizes non-linear chaotic systems to produce secure cryptographic keys. This module is divided into two sub-modules: Key generation using the De Jong Fractal Map, and key generation using the Van der Pol Oscillator.

\subsubsection{De Jong Fractal Map for Keys \( k_1 \) and \( k_2 \)}
\begin{figure*}[htbp]
    \centering
    
    \begin{subfigure}{0.18\textwidth}
        \centering
        \includegraphics[width=\linewidth]{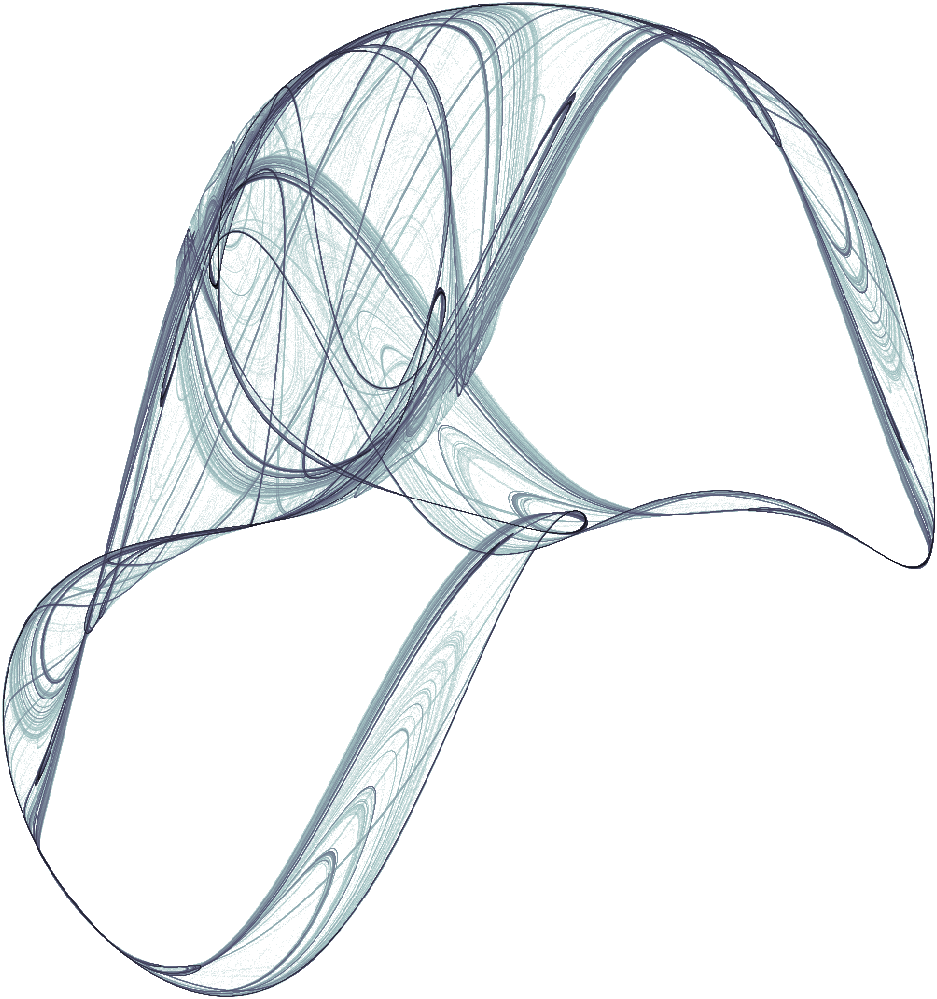}
        \caption{}
    \end{subfigure}%
    \hfill
    \begin{subfigure}{0.18\textwidth}
        \centering
        \includegraphics[width=\linewidth]{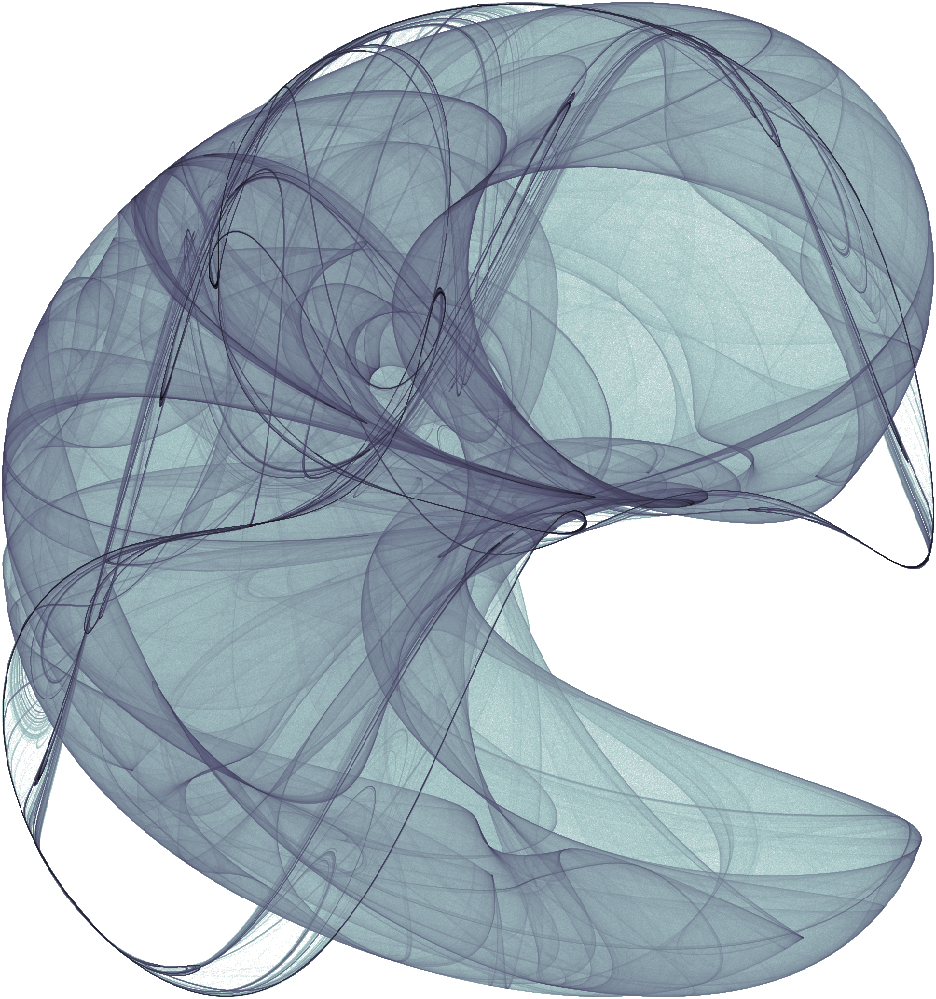}
        \caption{}
    \end{subfigure}%
    \hfill
    \begin{subfigure}{0.18\textwidth}
        \centering
        \includegraphics[width=\linewidth]{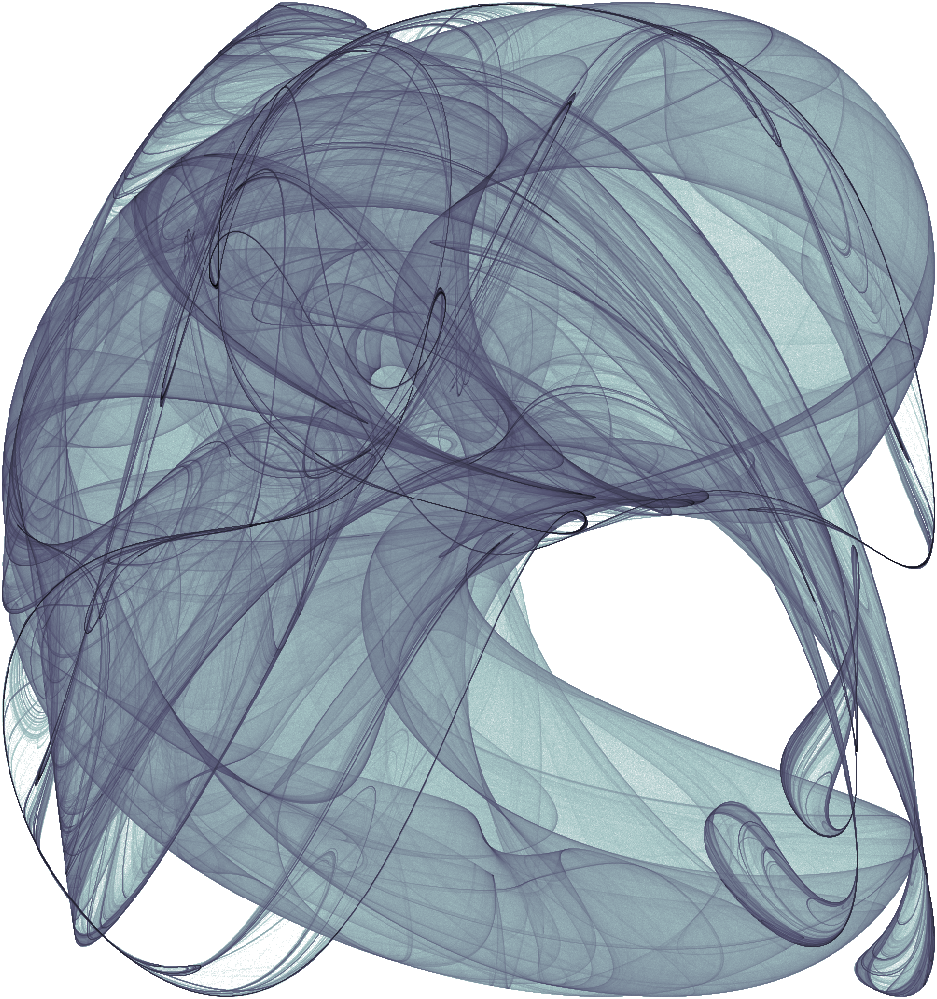}
        \caption{}
    \end{subfigure}%
    \hfill
    \begin{subfigure}{0.18\textwidth}
        \centering
        \includegraphics[width=\linewidth]{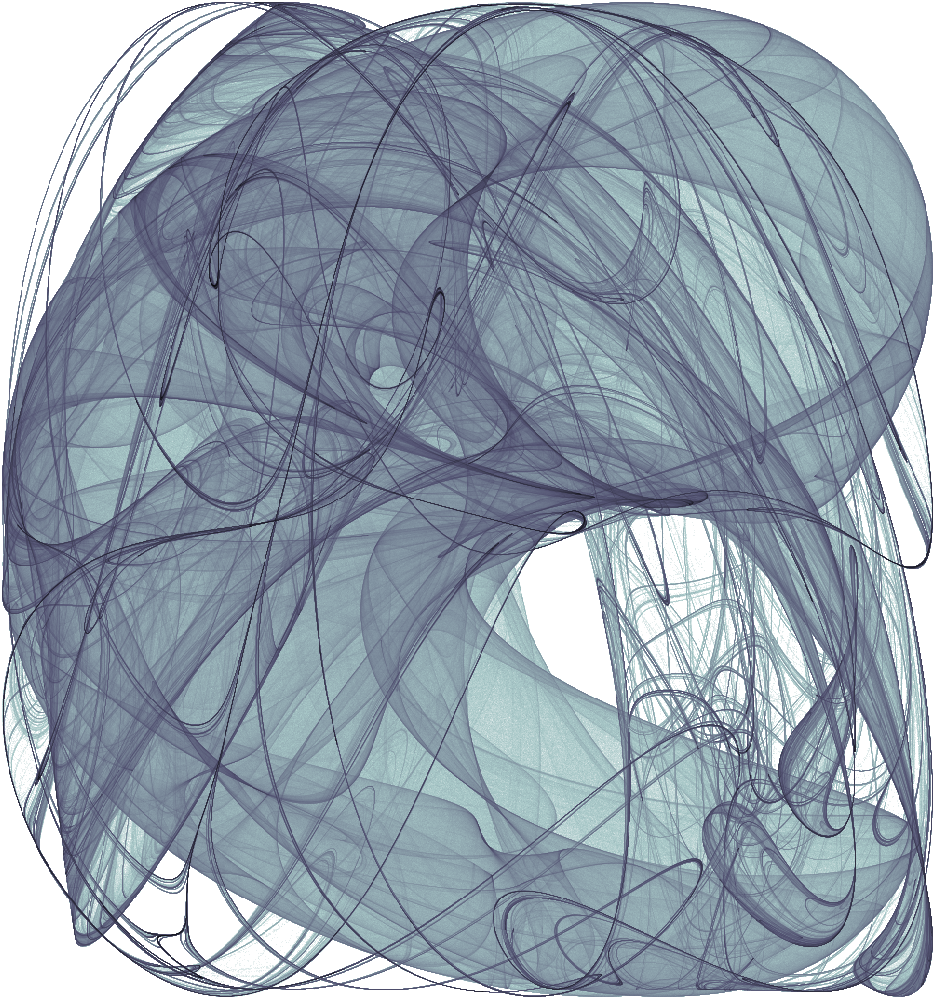}
        \caption{}
    \end{subfigure}%

    \caption{Parameter space exploration of the De Jong Fractal map for key generation: (a) a = 1.641, b = 1.902, c = 0.316, d= 1.525; (b) a = 1.4, b = -2.3, c = 2.4, d = -2.1; (c) a=2.01, b=-2.53, c=1.61, d=-0.33; (d) a=-2.7, b=-0.09, c=-0.86, d=-2.2.}
    \label{fig:fig2}
\end{figure*}

The De Jong Fractal Map is a strange chaotic attractor named after its creator, Peter de Jong \cite{dewdney1987probing} and is defined by the following set of parametric equations:
\begin{align}
x_{n+1} &= \sin(a \cdot y_n) - \cos(b \cdot x_n) \\
y_{n+1} &= \sin(c \cdot x_n) - \cos(d \cdot y_n)
\end{align}
Where \( x_{n+1} \) and \( y_{n+1} \) are the new coordinates, \( x_n \) and \( y_n \) are the previous coordinates, and the parameters \( a \), \( b \), \( c \), and \( d \) can be varied to produce different chaotic patterns shown in Figure \ref{fig:fig2}.

The proposed encryption scheme uses this map to generate two cryptographic keys: \(k_1\) for the chaotic operation selection in the transformative substitution module, and \(k_2\) for the key controlled Operation1—the modulus addition operation. The key generation process consists of the following steps:\\

\noindent\textbf{a. Setting Control Parameters and Initial Conditions:}
First step is to define the control parameters for the De Jong map.The utlized parameters are:

\noindent Control parameters:
\( \alpha = 1.4 \) \hfill \( \beta = 1.56 \) \hfill \( \gamma = 1.4 \) \hfill \( \delta = -6.56 \) \\
\( \epsilon = -1.6 \) \( \zeta = -0.2 \) \( \eta = 2.0 \) \( \lambda = 1.0 \)

Moroever, the initial values for \( x \) and \( y \) are set to 0. The number of iterations determines the length of the key. Here, the size of \(k_1\) is \( 256 \times 256 \).\\

\noindent\textbf{b. Generating the De Jong Map:}
The De Jong map is generated using the following equations. This step iteratively computes the values of x and y and results in a matrix called \(DeJong-matrix\).
\begin{equation}
\begin{aligned}
x_i &= a \sin(y_{\text{prev}} \times \beta) - c \cos(x_{\text{prev}} \times \delta) \\
y_i &= e \sin(x_{\text{prev}} \times \zeta) - g \cos(y_{\text{prev}} \times \lambda)
\end{aligned}
\end{equation}

\noindent\textbf{c. Generating Key \(k_1\):}
The first key, \( k1 \), is generated by taking the modulus \( 3 \) of the \(DeJong-matrix\), which is a \( 256 \times 256 \) matrix. The modulus operation given below ensures that each element of \( k1 \) will have a value of either 0, 1, or 2.
\[
k1_{i,j} = \textit{DeJong-matrix}_{i,j} \mod 3
\]

\noindent\textbf{d. Generating Key \(k_2\):}
For \( k2 \), we first compute a hash value from the \textit{DeJong-matrix} as follows:
\[
\text{hash} = \sum_{i=1}^{256} \sum_{j=1}^{256} \textit{DeJong-matrix}_{i,j}
\]
Using hash value, \( k2 \) is generated by extracting its 8 least significant bits by using:
\[
k2_{\text{bit}} = \text{bitget}(\text{hash\_value}, \text{bit})
\]

Where \text{bit} ranges from 1 to 8, and the \text{bitget} function retrieves the value of the specified bit from the hash value. The resulting \( k2 \) is an 8-bit binary key.

\subsubsection{Van Der Pol Oscillator for RNA Key \(k_3\)}

\begin{figure*}[htbp]
    \centering
    \begin{subfigure}[b]{0.18\textwidth}
        \includegraphics[width=\textwidth]{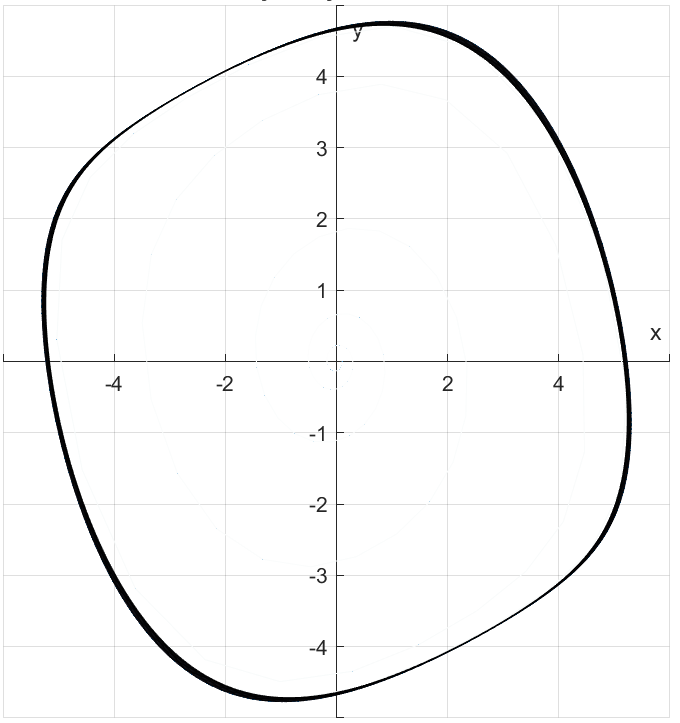}
        \caption{}
        \label{fig:a}
    \end{subfigure}
    \hfill
    \begin{subfigure}[b]{0.18\textwidth}
        \includegraphics[width=\textwidth]{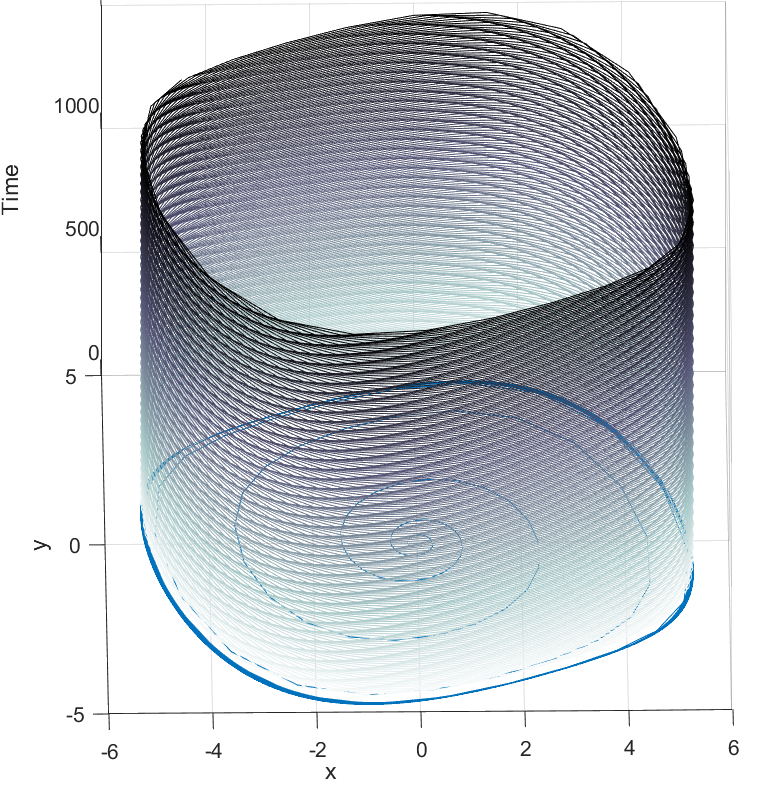}
        \caption{}
        \label{fig:b}
    \end{subfigure}
    \hfill
    \begin{subfigure}[b]{0.18\textwidth}
        \includegraphics[width=\textwidth]{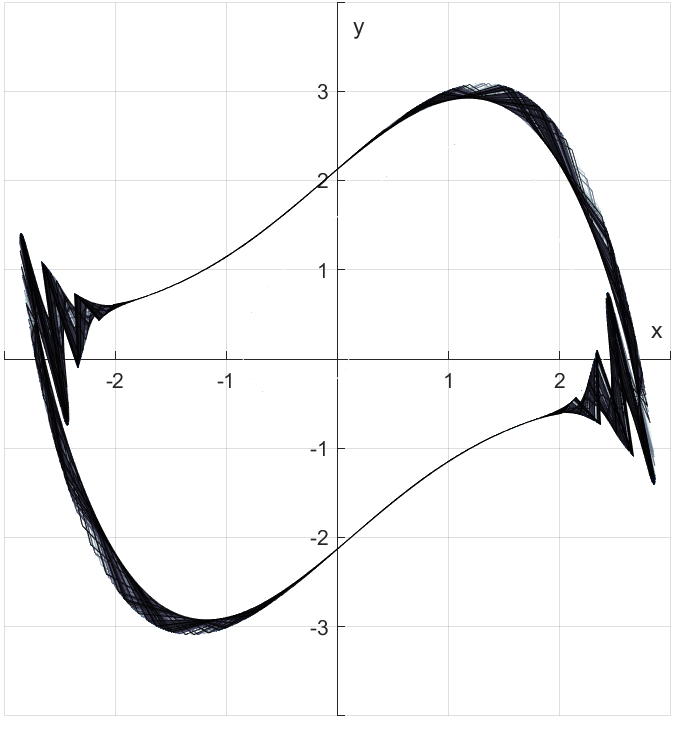}
        \caption{}
        \label{fig:c}
    \end{subfigure}
    \hfill
    \begin{subfigure}[b]{0.18\textwidth}
        \includegraphics[width=\textwidth]{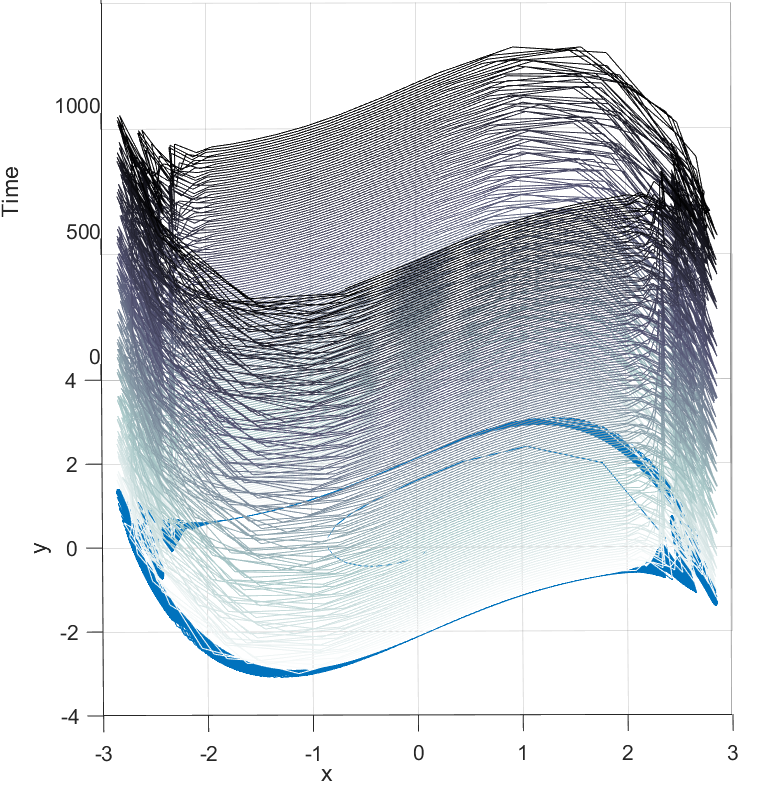}
        \caption{}
        \label{fig:d}
    \end{subfigure}
    
    \caption{The parameter space exploration for Van Der Oscillator for: (a,b) \( h=0.3 \) and \( \mu=0.05 \); and (c,d) \( h=0.3 \) and \( \mu=0.5 \).}
    \label{fig:fig3}
\end{figure*}

The key used in the RNA encoding module is generated using the discretized Van der Pol oscillator. It is a non-linear second-order differential equation introduced by Balthasar van der Pol \cite{van1920theory} and  can be discretized as:
\begin{align}
\frac{x_{n+1} - x_n}{h} &= v_n \\
\frac{v_{n+1} - v_n}{h} &= \mu(1-x_n^2)v_n - x_n
\end{align}
where \( x_n \) and \( v_n \) are the approximations of \( x(t) \) and \( \frac{dx}{dt} \) at time \( t = nh \), respectively. The parameters \( h \) and \( \mu \) describe its chaotic as visualized in Figure \ref{fig:fig3}. Detailed steps of key generation are:
\vskip0.5pc
\noindent\textbf{a. Initial conditions and Time Span:}
The first step is to establish the initial conditions of the oscillator, denoted as $y_0$, and a specified time span, $t_{\text{span}}$.
\vskip0.5pc
\noindent\textbf{b. Define and solve the Differential Equations:}
Define the set of non-linear differential equations as follows and solve them over the time span $t_{\text{span}}$. The solutions yield two sets of time series data, which are then stored in matrix \(Y\).
\begin{align}
\dot{y}_1 & = y_2 \\
\dot{y}_2 & = (1 - y_1^2) \times y_2 - y_1
\end{align}
Where \(\dot{y}_1\) and \(\dot{y}_2\) represent the time derivatives.
\vskip0.5pc
\noindent\textbf{c. Conversion to Indices and Permutation:}
The normalized sequence is then scaled and rounded to produce a set of indices. This transformation is done as:
\begin{equation}
\text{indices} = \text{round}(\text{normalized\_sequence} \times 64) + 1
\end{equation}
This step ensures that the indices are integers ranging from 1 to 65. For each value of \( i \) from 1 to the length of the indices, final index \( \text{idx} \) is calculated as:
\begin{equation}
\text{idx} = \mod(i + \text{indices}(i) - 1, 65) + 1
\end{equation}

\begin{algorithm} [H]
\small
\caption{Key Generation \((k_1, k_2)\) Using the De Jong Map}
\begin{algorithmic}[1]
\REQUIRE Control parameters \( \alpha, \beta, \gamma, \delta, \epsilon, \zeta, \eta, \lambda \) and initial conditions \( x_0, y_0 \)
\ENSURE \( k_1 \)  (256x256 matrix), \( k_2 \)  (8-bit binary key)

\STATE \textbf{function} GENERATE\_KEYS(\( \alpha, \beta, \gamma, \delta, \epsilon, \zeta, \eta, \lambda, x_0, y_0 \))
\STATE Initialize \( x_{\text{values}}[1] \leftarrow x_0 \), \( y_{\text{values}}[1] \leftarrow y_0 \)
\FOR{\( i = 2 \) to \( 256 \times 256 \)}
    \STATE \( x_{\text{prev}} \leftarrow x_{\text{values}}[i - 1] \)
    \STATE \( y_{\text{prev}} \leftarrow y_{\text{values}}[i - 1] \)
    \STATE \( x_{\text{values}}[i] \leftarrow \alpha \times \sin(y_{\text{prev}} \times \beta) - \gamma \times \cos(x_{\text{prev}} \times \delta) \)
    \STATE \( y_{\text{values}}[i] \leftarrow \epsilon \times \sin(x_{\text{prev}} \times \zeta) - \eta \times \cos(y_{\text{prev}} \times \lambda) \)
\ENDFOR
\STATE Normalize and convert \( x_{\text{values}} \) and \( y_{\text{values}} \) to 8-bit integers
\STATE \( \text{dejong\_matrix} \leftarrow \) Reshape of \( x_{\text{values}} \)
\STATE \( k_1 \leftarrow \text{dejong\_matrix} \mod 3 \)
\STATE Compute hash value from dejong\_matrix
\STATE \( k_2 \leftarrow \) 8 least significant bits of the hash value
\STATE \textbf{return} \( k_1, k_2 \)
\STATE \textbf{end function}
\end{algorithmic}
\end{algorithm}

\begin{algorithm} [H]
\small
\caption{Key Generation \((k_3)\) Using Van Der Pol Oscillator}
\begin{algorithmic}[1]
\REQUIRE Initial conditions \( y_0 \), Time span \( t_{\text{span}} \)
\ENSURE Permutation Key \( k_3 \) (unique numbers ranging from 0 to 64)

\STATE \textbf{function} VANDERPOL\_CPRNG(\( y_0, t_{\text{span}} \)) \text{\textbf{returns}} \text{sequence}

\STATE Define the Van der Pol differential equations:
\STATE \hspace{\algorithmicindent} \( \dot{y}_1 \leftarrow y_2 \)
\STATE \hspace{\algorithmicindent} \( \dot{y}_2 \leftarrow (1 - y_1^2) \times y_2 - y_1 \)
\STATE \( Y \leftarrow \text{results} \)
\STATE Normalize the sequence:
\STATE \hspace{\algorithmicindent} \( \text{normalized\_sequence} \leftarrow \frac{Y(:,1) - \min(Y(:,1))}{\max(Y(:,1)) - \min(Y(:,1))} \)
\STATE Convert the sequence to indices:
\STATE \hspace{\algorithmicindent} \( \text{indices} \leftarrow \text{round}(\text{normalized\_sequence} \times 64) + 1 \)
\STATE Initialize a list:
\STATE \hspace{\algorithmicindent} \( \text{numbers} \leftarrow [0,1,2,\dots,64] \)
\FOR{\( i = 1 \) to \( \text{length(indices)} \)}
\STATE \( \text{idx} \leftarrow \text{mod}(i + \text{indices}(i) - 1, 65) + 1 \)
\IF{\( i \leq 65 \) and \( \text{idx} \leq 65 \)}
\STATE Swap the values of \( \text{numbers}(i) \) and \( \text{numbers}(\text{idx}) \)
\ENDIF
\ENDFOR
\STATE Extract key \( k_3 \)
\STATE \textbf{end function}
\end{algorithmic}
\end{algorithm}

If both \( i \) and \( \text{idx} \) are within the range of 1 to 65, the values at positions \( i \) and \( \text{idx} \) in the numbers list are swapped. This process effectively permutes the numbers list in a pseudo-random fashion guided by the dynamics of the Van der Pol oscillator.\\

\noindent\textbf{d. Extract Key \(k_3\):}
The final key sequence is derived by extracting the first 64 numbers from the permuted numbers list.

\subsection{The RNA Encoding Module}

This paper presents a novel two-base RNA encoding method, in which the image is converted into a simulated RNA sequence and then this sequence is shuffled using the chaotic map. The block diagram given in Figure \ref{fig:fig5} depicts the RNA encoding process, which is explained as follows:\\

\noindent\textbf{1. Defining Two-Base Combinations:}
RNA is made up of four bases: Adenine (A), Uracil (U), Cytosine (C), and	Guanine (G). We use 16 two-base combinations, i.e., there are 4 bases, the possible combinations become 4x4=16. Fig. \ref{fig:RNA-table} shows the two-base combination rule used in this paper.\\

\noindent\textbf{2. Encoding Pixels Using RNA rules:}
In this step the original image is converted into an RNA strand-like sequence. The pixel values are mapped to 16 two-base combinations given in Figure \ref{fig:RNA-table}.

\begin{enumerate}[(a)]
    \item The pixel value is divided by 16 and floored, resulting in a number between 0 and 15. This number is then used as an index to select and map the 256 possible pixel values to one of the 16 RNA base combinations as per the RNA Rule given in Fig. \ref{fig:RNA-table}. Let's take a \(4 \times 4\) image for the ease of understanding:
{\footnotesize
\[
\text{plaintext} = \begin{pmatrix}
255 & 238 & 187 & 170 \\
221 & 204 & 153 & 136 \\
119 & 102 & 51 & 34 \\
85 & 68 & 17 & 0 \\
\end{pmatrix}
\]
}

\begin{figure}[H]
    \centering
    \includegraphics[width=0.48\textwidth]{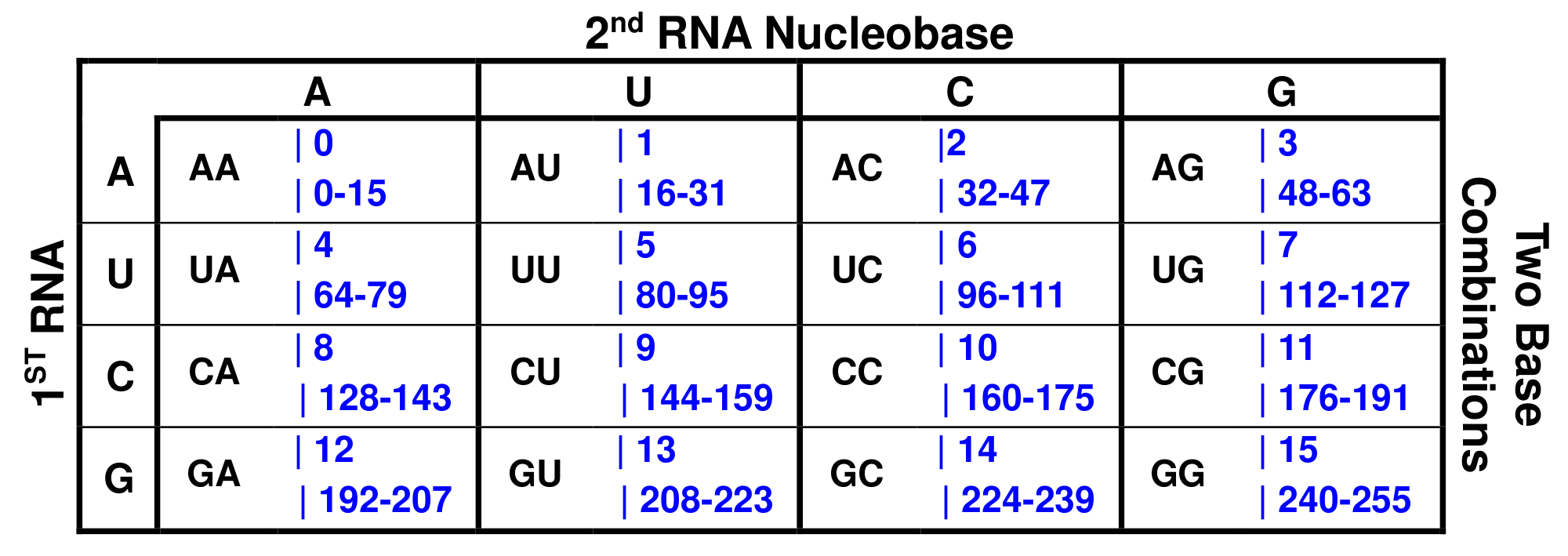}
    \caption{RNA two-base combination rule}
    \label{fig:RNA-table}
\end{figure}

    After mapping, the RNA sequence matrix is generated as:
 {\footnotesize
\begin{equation*}
    \text{RNA sequence matrix} = \begin{pmatrix}
    GG & GC & CG & CC \\
    GU & GA & CU & CA \\
    UG & UC & AG & AC \\
    UU & UA & AU & AA \\
    \end{pmatrix}
\end{equation*}
}

\item The RNA sequence matrix is then appended to get the RNA sequence:
    {
\footnotesize
\[
\begin{aligned}
\text{RNA Sequence} &= \begin{bmatrix}
GG||GC & GU||GA & CG||CC & CU||CA \\
UG||UC & UU||UA & AG||AC & AU||AA \\
\end{bmatrix}
\end{aligned}
\]
}

   {
\small
\[
\begin{aligned}
\text{RNA Sequence} &= [GGGCGUGACGCCCUCA... \\
  & ...UGUCUUUAAGACAUAA]
\end{aligned}
\]
}
\end{enumerate}

\noindent\textbf{3. Block Division and Chaotic Permutation:}
The RNA sequence is divided into blocks of 4 bases each. For the \(4 \times 4\) image, the RNA sequence has 32 bases. This will result in \( \frac{32}{4} = 8 \) blocks.
   {
\small
\[
\begin{aligned}
\text{RNA Sequence Blocks} = [\texttt{GGGC}, \texttt{GUGA}, \texttt{CGCC}, \texttt{CUCA},\\ \texttt{UGUC}, \texttt{UUUA}, \texttt{AGAC}, \texttt{AUAA}]
\end{aligned}
\]
}
The blocks are then permuted with the help of a permutation key. This key is generated using the De Jong Fractal map. This key indicates the new order of RNA blocks.The permuted RNA sequence is:
  {
\small
\[
\begin{aligned}
\text{Permuted RNA Sequence} &= [UGUCCGCCAUAAGGGC... \\
  & ...AGACCUCAGUGAUUUA]
\end{aligned}
\]
}

\noindent\textbf{4. RNA Encrypted Image:}
The permuted RNA bases correspond to the permuted pixels and hence, generate the RNA encrypted image as:
{\footnotesize
\begin{equation*}
    \text{RNA Encrypted Matrix} = \begin{bmatrix}
    119 & 102 & 17 & 0 \\
    187 & 170 & 255 & 238 \\
    51 & 34 & 221 & 204 \\
    153 & 136 & 85 & 68 \\
    \end{bmatrix}
\end{equation*}
}

\subsection{The Novel Transformative Substitution Module}
This paper introduces a unique transformative substitution method given in Fig. \ref{fig:fig6}. In this method, the s-box value is transformed before it replaces the pixel in the image. The transformation applied to each chosen value varies each time, based on the selected operation. In this regard, three cryptographic operations tailored for image encryption are proposed in this paper.

\begin{figure*}[t]
    \centering
    \includegraphics[width=0.8\textwidth]{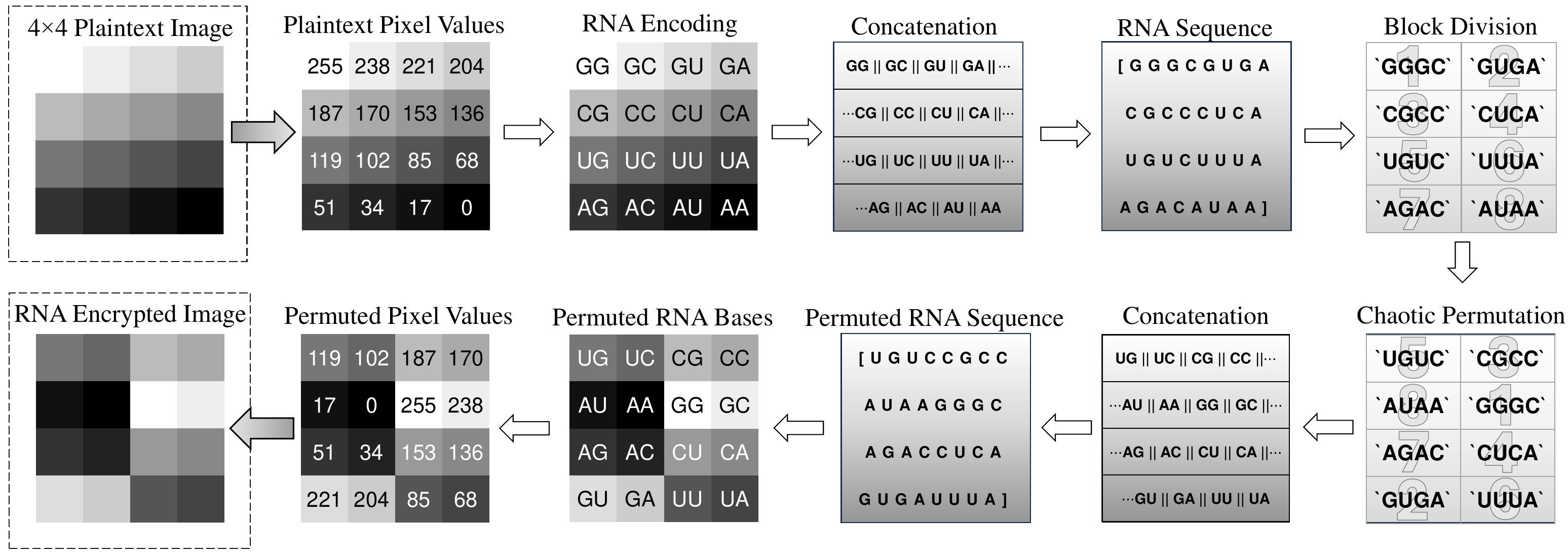}
    \caption{RNA encoding technique for a \(4 \times 4\) grayscale image}
    \label{fig:fig5}
\end{figure*}

    \begin{figure*}[!t]
    \centering
    \includegraphics[width=0.8\textwidth]{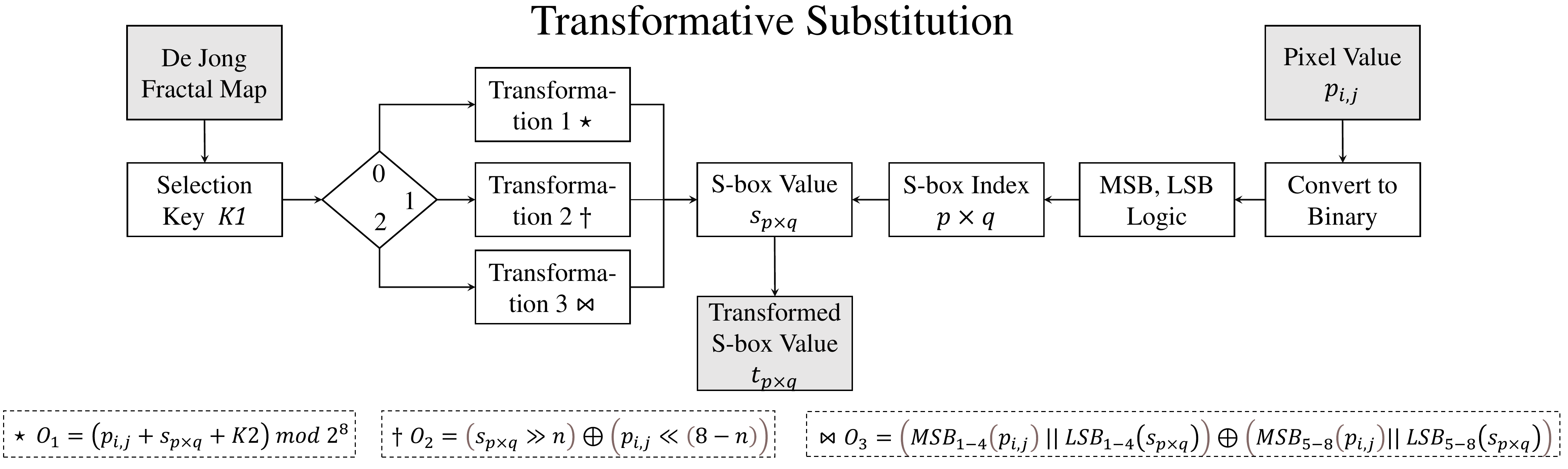}
    \caption{The novel transformative substitution module with cryptographic operations}
    \label{fig:fig6}
\end{figure*}

\subsubsection{Specialized Cryptographic Operations for Image Encryption}

The following three cryptographic operations are presented in this paper tailored for image encryption.\\

\textbullet{\textbf{ Operation 1:}} \( \bm{O_1 = (p_{i,j} + s_{p \times q} + K_2) \mod 2^8} \)\\
This is a key dependent modulus addition operation. The key \( K_2 \) is generated by the De Jong Fractal Map. The use of the modulus operation adds a layer of cyclic complexity, making the output harder to predict. Here, \( p_{i,j} \) represents the pixel value, and \( s_{p \times q} \) represents the selected S-box value.
\vskip.5pc
\textbullet{\textbf{ Operation 2:}}  \( \bm{O_2 = (s_{p \times q} \gg n) \oplus (p_{i,j} \ll (8-n))} \)
\\
This operation employs bitwise shifting and XOR functions to randomize the encrypted values, thereby increasing resistance to linear analysis techniques.
\vskip.5pc
\textbullet \textbf{ Operation 3:} 
\( \bm{O_3 = (\text{MSB}_{1-4}(p_{i,j}) \, || \, \text{LSB}_{1-4}(s_{p \times q}))} \\ 
\bm{\oplus (\text{MSB}_{5-8}(p_{i,j}) \, || \, \text{LSB}_{5-8}(s_{p \times q}))} \)\\
This operation involves the concatenation and XORing of the Most Significant Bits (MSB) and Least Significant Bits (LSB) of the s-box value and the pixel value, adding unpredictability and complexity between the original and transformed pixels.

\section{Results and Discussion}

\subsection{Statistical Security Analysis}
Entropy gauges randomness with an ideal value of 8 for an 8-bit image, while GLCM (Gray-Level Co-occurrence Matrix) parameters like correlation, contrast, homogeneity, and energy assess texture properties and spatial relationships. The results in Table \ref{table:your_label} confirm that our encryption scheme yields values that closely align with these ideals, showcasing its strong statistical security.

\subsection{Histogram Analysis}
\begin{figure*}[t!]
    \centering
    \begin{subfigure}[b]{0.17\textwidth}
        \includegraphics[width=\textwidth]{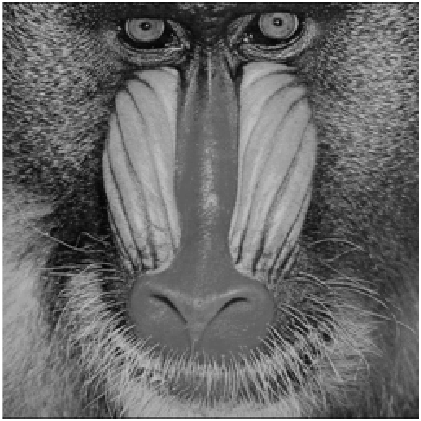}
        \caption{}
        \label{fig:a}
    \end{subfigure}
    \hfill
    \begin{subfigure}[b]{0.20\textwidth}
        \includegraphics[width=\textwidth]{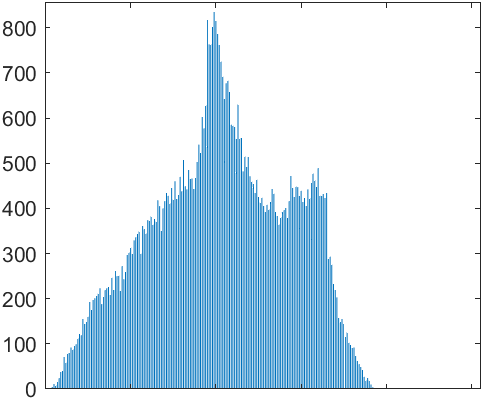}
        \caption{}
        \label{fig:b}
    \end{subfigure}
    \hfill
    \begin{subfigure}[b]{0.17\textwidth}
        \includegraphics[width=\textwidth]{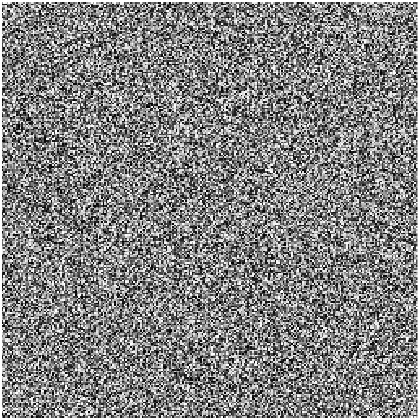}
        \caption{}
        \label{fig:c}
    \end{subfigure}
    \hfill
    \begin{subfigure}[b]{0.20\textwidth}
        \includegraphics[width=\textwidth]{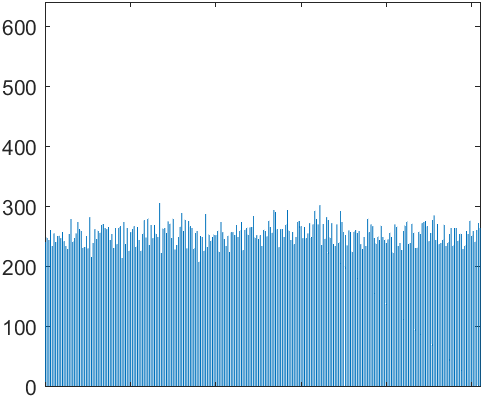}
        \caption{}
        \label{fig:d}
    \end{subfigure}

    \begin{subfigure}[b]{0.17\textwidth}
        \includegraphics[width=\textwidth]{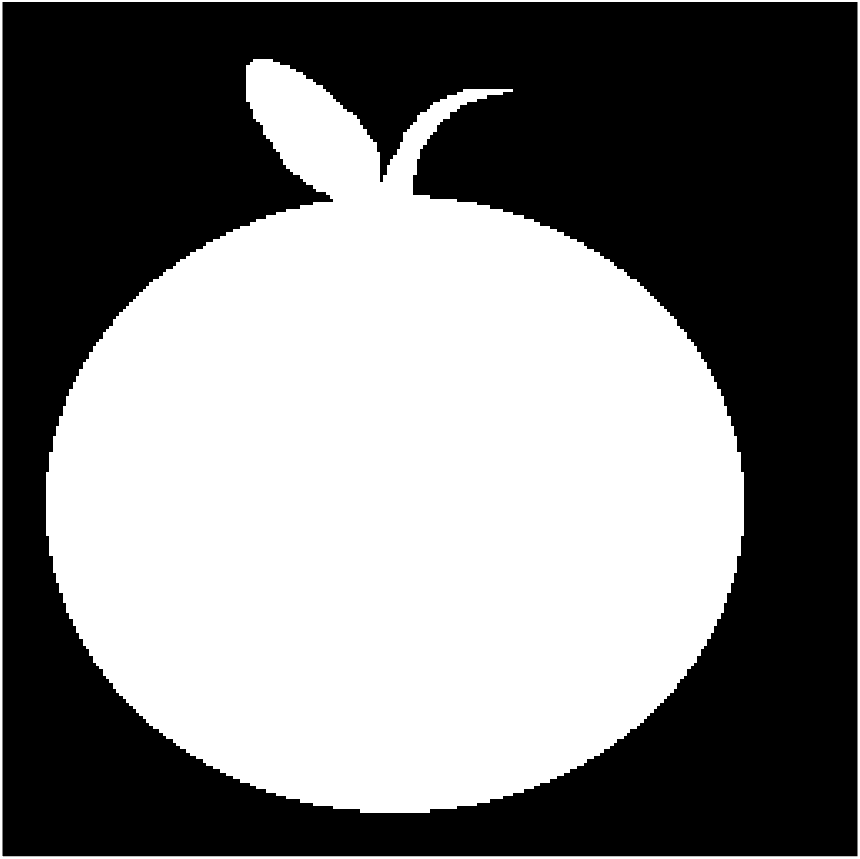}
        \caption{}
        \label{fig:e}
    \end{subfigure}
    \hfill
    \begin{subfigure}[b]{0.20\textwidth}
        \includegraphics[width=\textwidth]{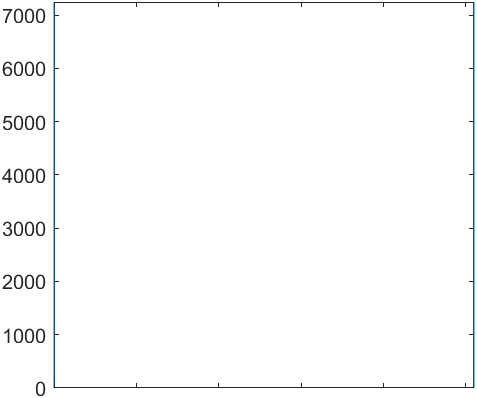}
        \caption{}
        \label{fig:f}
    \end{subfigure}
    \hfill
    \begin{subfigure}[b]{0.17\textwidth}
        \includegraphics[width=\textwidth]{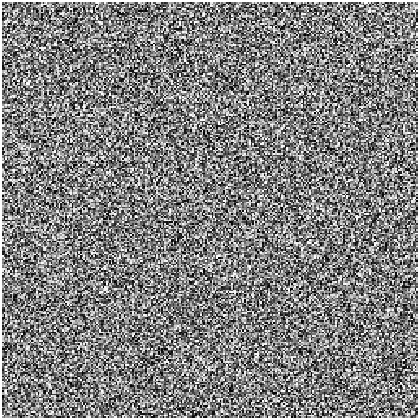}
        \caption{}
        \label{fig:g}
    \end{subfigure}
    \hfill
    \begin{subfigure}[b]{0.20\textwidth}
        \includegraphics[width=\textwidth]{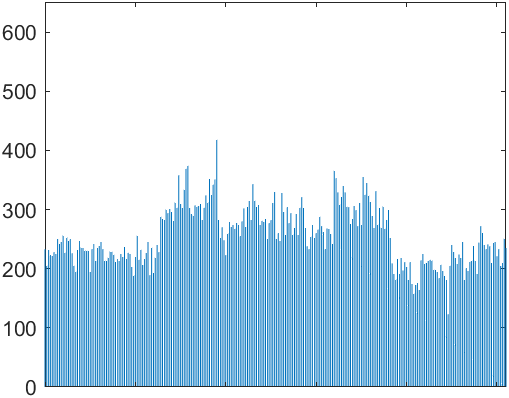}
        \caption{}
        \label{fig:h}
    \end{subfigure}

    \caption{Histogram analysis of the: (a-d) Baboon image; (e-h) Binary Apple image.}
    \label{fig:fig7}
\end{figure*}

Histogram analysis evaluates the distribution of pixel values in an encrypted image. A uniform histogram indicates that pixel values are evenly spread, reducing the chances of predictability and susceptibility to attacks. It is evident from Figure \ref{fig:fig7} that the histograms of our proposed scheme display a uniform distribution, signifying the effectiveness of our encryption method. 

\subsection{Correlation Analysis}
\begin{figure*}
    \centering
    \begin{subfigure}[b]{0.2\textwidth}
        \includegraphics[width=\textwidth]{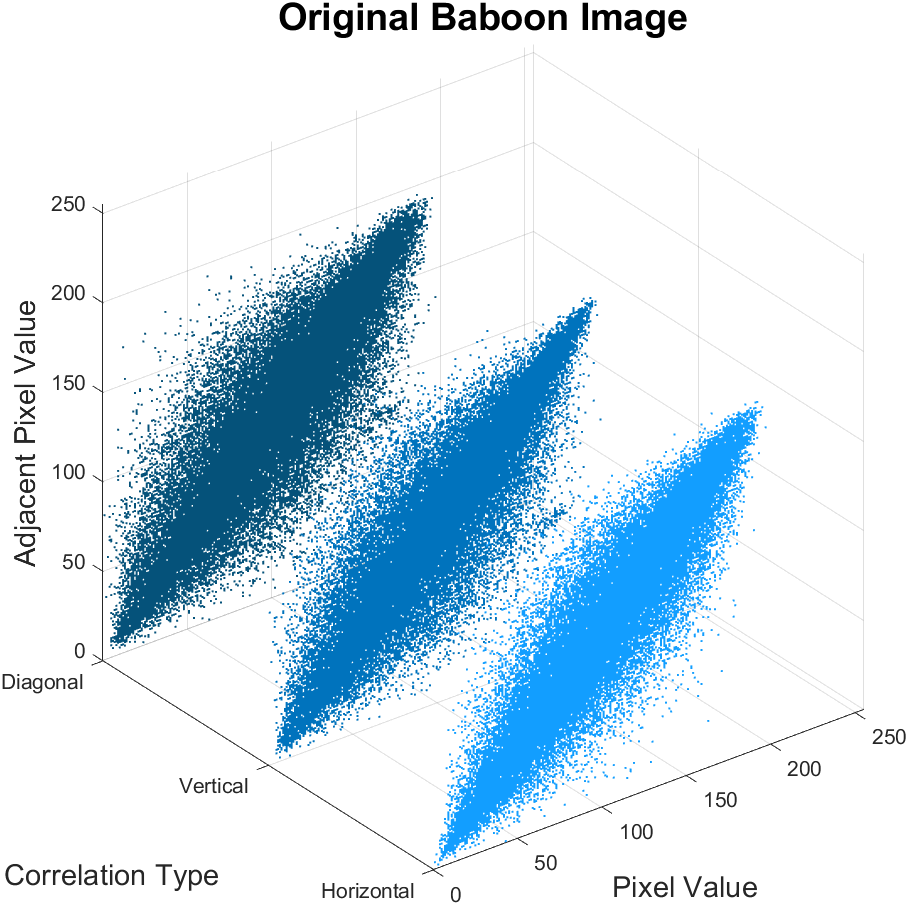}
        \caption{}
        \label{fig:a}
    \end{subfigure}
    \hfill
    \begin{subfigure}[b]{0.2\textwidth}
        \includegraphics[width=\textwidth]{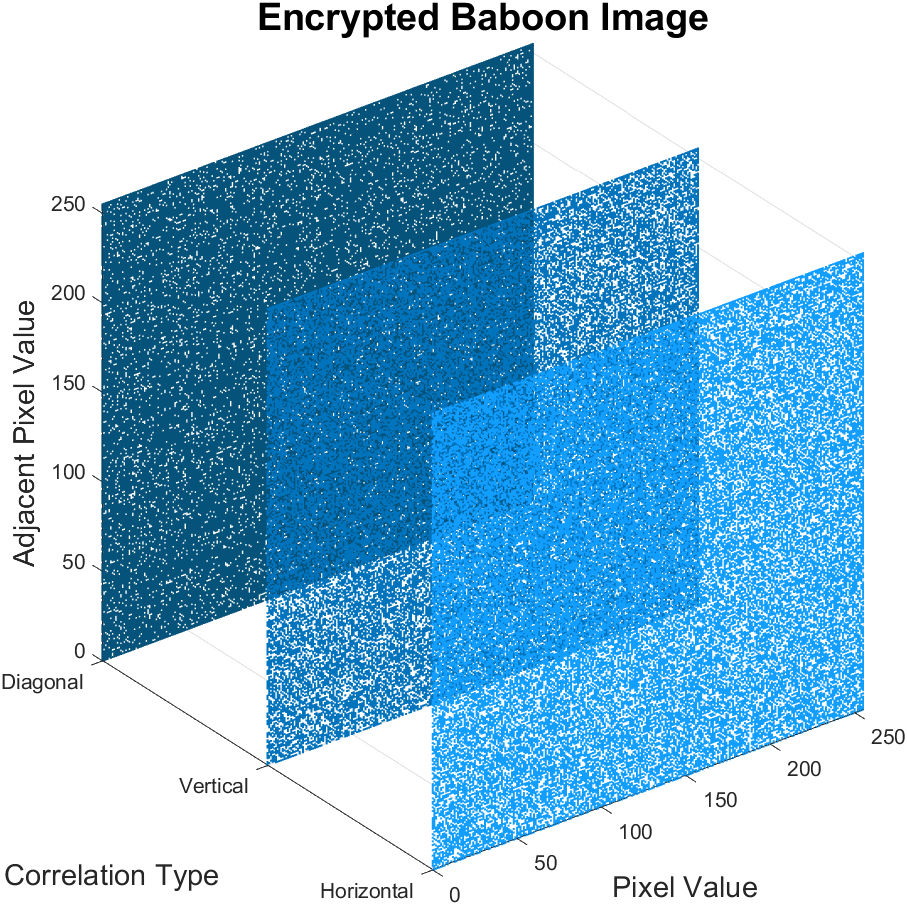}
        \caption{}
        \label{fig:b}
    \end{subfigure}
    \hfill
    \begin{subfigure}[b]{0.2\textwidth}
        \includegraphics[width=\textwidth]{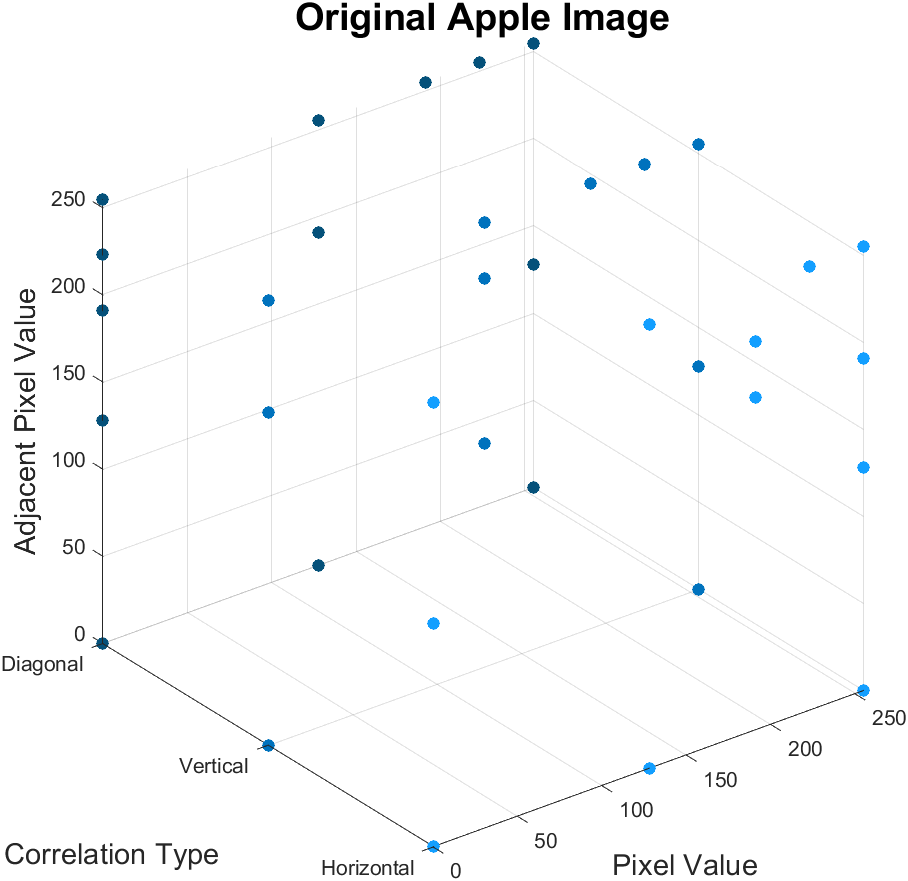}
        \caption{}
        \label{fig:c}
    \end{subfigure}
    \hfill
    \begin{subfigure}[b]{0.2\textwidth}
        \includegraphics[width=\textwidth]{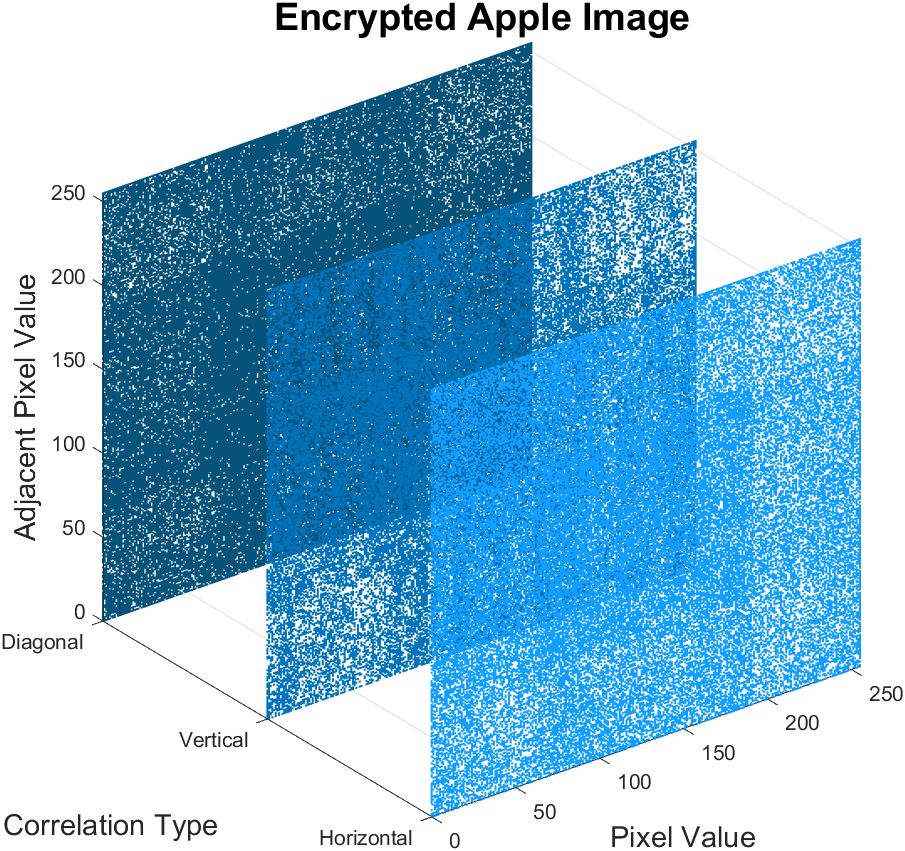}
        \caption{}
        \label{fig:d}
    \end{subfigure}
    
    \caption{Correlation analysis ofthe: (a) Plaintext Baboon; (b) Encrypted Baboon; (c) Plaintext Binary Apple; (d) Encrypted Binary Apple.}
    \label{fig:fig8}
\end{figure*}

Correlation analysis evaluates the statistical dependency between adjacent pixels in an image. It can be observed in Fig. \ref{fig:fig8} that the correlation coefficients for all three directions have been effectively scrambled, emphasizing the robustness of our encryption scheme.
\begin{table}[H]
    \centering
    \caption{Statistical Security Parameters of Encrypted Images}
    \label{table:your_label}
    \renewcommand{\makecell}[2][tl]{\begin{tabular}[#1]{@{}c@{}}#2\end{tabular}}
    \begin{tabularx}{\linewidth}{lXXXXX}
        \hline
        {\footnotesize \textbf{Test Images}} & {\footnotesize \textbf{Entropy}} & {\footnotesize \makecell{\textbf{Correl-} \\ \textbf{ation}}} & {\footnotesize \makecell{\textbf{Cont-} \\ \textbf{rast}}} & {\footnotesize \textbf{Energy}} & {\footnotesize \makecell{\textbf{Homo-} \\ \textbf{geneity}}} \\
        \hline
        Baboon Image & 7.996 & 0.00006 & 10.4 & 0.01 & 0.38 \\
        Apple Image  & 7.97  & 0.001   & 9.48 & 0.01 & 0.39 \\
        \hline
    \end{tabularx}
\end{table}

\section{Conclusion}
A novel image encryption scheme, RNA-TransCrypt, tailored for Internet of Things (IoT) devices was presented in response to the growing cybersecurity threats to image data. This scheme is not only secure but also caters the resource constrained nature of the IoT devices, focusing on a lightweight architecture. RNA-TransCrypt combines the biocryptographic capabilities of RNA encoding and the inherent unpredictability of chaos theory. The scheme’s innovation encompasses a unique two-base RNA encoding method for diffusion/pixel scrambling, a transformative substitution technique for efficient confusion/pixel value replacement, and three cryptographic operations specialized for image encryption. These operations introduced additional layers of security without imposing computational strains. The results demonstrated the proposed scheme's efficacy, as evidenced by an entropy of 7.997 and a correlation of 0.0006 results, confirming the successful scrambling and concealing of visual information.

\bibliographystyle{iet}
\bibliography{main}

\end{multicols}

\end{document}